\documentclass[notitlepage,onecolumn]{article}
\usepackage{sw20lart}

\input{tcilatex}
\begin{document}

\title{Consistency and Linearity in Quantum Theory}
\author{Ariel Caticha \\
{\small Department of Physics, University at Albany-SUNY, Albany, NY 12222}}
\date{}
\maketitle

\begin{abstract}
Quantum theory is formulated as the uniquely consistent way to manipulate
probability amplitudes. The crucial ingredient is a consistency constraint:
if the amplitude of a quantum process can be computed in two different ways,
the two answers must agree. The constraint is expressed in the form of
functional equations the solution of which leads to the usual sum and
product rules for amplitudes. An immediate consequence is that the
Schr\"{o}dinger equation must be linear: non-linear variants of quantum
mechanics violate the requirement of consistency.\newline
PACS: 03.65.Bz, 03.65.Ca.
\end{abstract}

The formalism of quantum mechanics is very robust; modifications are
invariably unsuccessful. A number of non-linear alternative theories have
been proposed,$^{\cite{bialynicki76}}$ experimentally tested,$^{\cite
{shull80}}$ and promptly discarded, but the question of whether new
non-linear variants might prove successful is not settled. In fact, interest
in the subject remains in spite of, or perhaps, partly because of the
possibility that these theories might allow super-luminal communication.$^{%
\cite{gisin90}}$ \ 

In this paper we propose an explanation for the robustness of quantum
mechanics. The basic argument is analogous to one used by R. T. Cox in the
very different context of classical probability theory. He showed that once
degrees of probability are represented by real numbers there is a unique set
of rules for reasoning under conditions of insufficient information.$^{\cite
{cox46}}$ His analysis not only legitimized the view of probability theory
as an extended form of logic, but also provided an explanation for its
uniqueness, for its inevitability.$^{\cite{jaynes83}}$

The crux of our argument is (like Cox's) a consistency requirement: if the
amplitude (i.e., the probability amplitude) for a quantum process can be
computed in two different ways, the two answers must agree. This requirement
is expressed in the form of functional equations, the solution of which
leads to the usual sum and product rules for amplitudes.$^{\cite{feynman65}}$
An immediate consequence is that the equation for time evolution, the
Schr\"{o}dinger equation, is necessarily linear. In other words, the usual
linear quantum theory emerges as the unique way to manipulate amplitudes
consistently.

To emphasize the simplicity of our approach we will focus on a simple
particle with no spin; its only attribute is its position. The
generalization to other more complex systems should in principle be
straightforward.

The first step is pragmatic: we identify the statements we are allowed to
make about the particle with the idealized experimental setups designed to
test them. As will become clear below this set of allowed propositions or
setups is more restricted than those used in other approaches based on
either quantum logic,$^{\cite{finkelstein63}\text{-}\cite{hooker79}}$ or
consistent histories$^{\cite{isham94}}$ or on Bayesian complex probabilities.%
$^{\cite{youssef91}}$

Suppose that what we can detect is the presence or absence of a particle in
a small region of space-time around an event $x=\left(\vec{x},t\right)$. The
simplest statement describing motion, ``the particle moves from $x_i$ to $%
x_f $,'' will be denoted by $[x_f,x_i]$. It can be tested by preparing a
particle at $x_i$ and placing a detector at $x_f$.

The more complex proposition ``the particle goes from $x_i$ to $x_1$ and
from there to $x_f$'' (we assume that $t_i<t_1<t_f$) will be denoted by $%
[x_f,x_1,x_i]$. To test it we need to introduce an idealized device, a
``filter'' which prevents any motion from $x_i$ to $x_f$ except via the
intermediate event $x_1$. This filter is some sort of obstacle or screen
that exists only at time $t_1$, blocking the particle everywhere in space
except for a small ``hole'' around the point $\vec{x}_1$.

We can also open two holes, one at $\vec{x}_1$ and another at $\vec{x}%
_1^{\prime }$ in our filter at $t_1$. This will allow testing whether ``the
particle goes from $x_i$ to $x_f$ via point $x_1$ or $x_1^{\prime }$,''
which we may write as $[x_f,(x_1,x_1^{\prime }),x_i]$.

The possibility of introducing many filters each with many holes leads to
allowed propositions or setups of the general form

\begin{equation}
a=[x_f,s_N,s_{N-1},\ldots ,s_2,s_1,x_i]\text{ ,}  \label{propdef}
\end{equation}
where $s_n=(x_n,x_n^{\prime },x_n^{\prime \prime },\ldots )$ is a filter at
time $t_n$, intermediate between $t_i$ and $t_f$, with holes at $\vec{x}_n,%
\vec{x}_n^{\prime },\vec{x}_n^{\prime \prime },\ldots $These propositions
involving a single initial and a single final event are special cases of
what are usually called histories.$^{\cite{isham94},\cite{footnote1}}$

Our second step starts from the observation that if two setups are related
in some way then information about one may be relevant to predictions about
the other. A relation of this kind arises when two setups $a$ and $b$ are
placed in immediate succession resulting in a third setup which will be
denoted by $ab$. This operation, which we will call $and$, cannot be used to
combine any two arbitrarily chosen propositions $a$ and $b$; it is necessary
that the destination point of the earlier setup coincide with the source
point of the later setup, otherwise the combined $ab$ is not an allowed
proposition. For example, \ 
\begin{equation}  \label{and1}
[x_f,s_N,\ldots,s_{n+1},x_n][x_n,s_{n-1},\ldots,s_1,x_i]=[x_f,s_N,%
\ldots,x_n,\ldots,s_1,x_i]\,.
\end{equation}
\newline
It is crucial that all $t_1,\ldots,t_{n-1}$ happen before $\,t_n$ and that $%
t_{n+1},\ldots,t_N$ happen after $t_n$, otherwise the two setups are not
consecutive.

Another useful relation arises when two setups $a^{\prime }$ and $a^{\prime
\prime }$ are identical except on one single filter where none of the holes
of $a^{\prime }$ overlap any of the holes of $a^{\prime \prime }$. We may
then form a third setup $a$, denoted by $a^{\prime }\vee a^{\prime \prime }$%
, which includes the holes of both $a^{\prime }$ and $a^{\prime \prime }$.
An example of this operation, which we will call $or$, is$\,$ 
\begin{equation}
\lbrack x_f,\ldots ,s_n^{\prime },\ldots ,x_i]\vee [x_f,\ldots ,s_n^{\prime
\prime },\ldots ,x_i]=[x_f,\ldots ,s_n,\ldots ,x_i]\text{ ,}  \label{or1}
\end{equation}
where

\[
s_n^{\prime }=(x_n^{\prime 1},x_n^{\prime 2},\ldots ,x_n^{\prime
j}),\,s_n^{\prime \prime }=(x_n^{\prime \prime 1},x_n^{\prime \prime
2},\ldots ,x_n^{\prime \prime k}), 
\]
and 
\[
s_n=(x_n^{\prime 1},\ldots ,x_n^{\prime j},x_n^{\prime \prime 1},\ldots
,x_n^{\prime \prime k}). 
\]
Again, notice that it is only for special choices of propositions $a$ and $b$
that this $or$ operation will result in an allowed proposition $a\vee b$.

The basic properties of $and$ and $or$ can be obtained from the principle
that two propositions are equal when they are tested by setups with the same
distribution of filters and holes. For example, the $or$ operation is
commutative, but $and$ is not 
\begin{equation}
a\vee b=b\vee a,\,\,\,\,\,\,\,ab\neq ba\text{ .}
\end{equation}
\newline
There is an asymmetry implicit in the idea of placing setups in succession:
one setup is the earlier one. If $ab$ is allowed, $ba$ is not. The next
properties concern associativity. Given three consecutive setups $a$, $b$
and $c$ we can write 
\begin{equation}  \label{assocand}
\left(ab\right)c=a\left(bc\right)\equiv abc\text{ ,}
\end{equation}
\newline
provided $ab$ and $bc$ are allowed (then $abc$ is allowed too). Similarly,
for the $or$ operation we have \ 
\begin{equation}  \label{assocor}
\left(a\vee b\right)\vee c=a\vee\left(b\vee c\right)\equiv a\vee b\vee c%
\text{ ,}
\end{equation}
\newline
provided $\left(a\vee b\right)$, $\,\left(b\vee c\right)$, $\,\left(a\vee
b\right)\vee c$ and $a\vee\left(b\vee c\right)$ are all allowed. Finally, $%
and$ and $or$ enjoy a measure of distributivity, \ 
\begin{equation}  \label{distrib}
a(b\vee c)=\left(ab\right)\vee\left(ac\right)\,\,\,\,\text{or}%
\,\,\,\,\,(b\vee c)a=\left(ba\right)\vee\left(ca\right)\text{ .}
\end{equation}
\newline
Which of these two holds, if any, depends again on whether the relevant
propositions are allowed.

The quantum $and$ and $or$ operations introduced here are not logical but
rather physical connectives; they represent our idealized ability to
construct more complex physical setups out of simpler ones. They differ from
their Boolean and quantum logic analogues in that they do not operate
between arbitrary propositions (see the paragraphs above eqs. (\ref{and1})
and (\ref{or1})). Furthermore, distributivity fails in quantum logics but
not in the present approach.

Next, we assume that a quantitative representation of the relations ($and/or$%
) between setups can be obtained by assigning to each setup $a$ a single
complex number $\phi(a)$. By a 'representation' we mean that the assignment
of $\phi$s is such that relations among setups translate into relations
among the corresponding complex numbers. But why should such a
representation exist? Why complex numbers? No answer here; this is the
mysterious feature of quantum theory. It seems that a single complex number
is sufficient to convey the physically relevant information about a setup.

To be specific, suppose that two setups $a$ and $a^{\prime }$ can be
combined into $a\vee a^{\prime }$. Then given the numbers $\phi \left(
a\right) $ and $\phi \left( a^{\prime }\right) $ and the fact that the
relevant relation is $or$ one should be able to calculate $\phi \left( a\vee
a^{\prime }\right) $: there must exist a function $S$ such that 
\begin{equation}
\phi \left( a\vee a^{\prime }\right) =S(\phi \left( a\right) ,\phi \left(
a^{\prime }\right) ),
\end{equation}
and that this same function $S$ apply to any other setups that are similarly
related. Thus, $S$ represents $or$.

The existence of $S$ constrains the assignment of $\phi $s. For example, the
number $\phi (a\vee a^{\prime }\vee a^{\prime \prime })$ can be calculated
either as $\phi \left( \left( a\vee a^{\prime }\right) \vee a^{\prime \prime
}\right) $ or as $\phi \left( a\vee \left( a^{\prime }\vee a^{\prime \prime
}\right) \right) $, and the two ways must agree, 
\begin{equation}
S(\phi \left( a\vee a^{\prime }\right) ,\phi \left( a^{\prime \prime
}\right) )=S(\phi \left( a\right) ,\phi \left( a^{\prime }\vee a^{\prime
\prime }\right) ).
\end{equation}
\newline
Further use of $S$ leads to the consistency constraint \ 
\begin{equation}
S(S\left( u,v\right) ,w)=S(u,S\left( v,w\right) ),  \label{constrs}
\end{equation}
\ where we have put $\phi \left( a\right) =u$, $\phi \left( a^{\prime
}\right) =v$, and $\phi \left( a^{\prime \prime }\right) =w$. One can check,
by substitution, that eq. (\ref{constrs}), is satisfied if 
\begin{equation}
S\left( u,v\right) =\xi ^{-1}(\xi \left( u\right) +\xi \left( v\right) ),
\end{equation}
or, 
\begin{equation}
\xi (S\left( u,v\right) )=\xi \left( u\right) +\xi \left( v\right) ,
\label{xis}
\end{equation}
where $\xi $ is an arbitrary function. Different choices of $\xi $ lead to
different, equally acceptable forms of $S$. The proof by Cox that eq. (\ref
{xis}) is the general solution in the case of real variables holds in the
complex case as well. Conversely,$^{\cite{cox46},\cite{caticha98}}$ if the
function $S$ exists then there must also exist another function $\xi $,
calculable from $S$, such that 
\[
\xi (\phi \left( a\vee a^{\prime }\right) )=\xi (\phi \left( a\right) )+\xi
(\phi \left( a^{\prime }\right) ). 
\]
This result shows that, instead of the original $\phi \left( a\right) $, an
equivalent and much more convenient assignment is the number $\xi \left(
\phi \left( a\right) \right) $, which we will denote by $\xi \left( a\right) 
$. In other words, the assignment of a number $\xi \left( a\right) $ to a
proposition $a$ can always be done so that $or$ is represented by a simple
sum rule, \ 
\begin{equation}
\xi \left( a\vee a^{\prime }\right) =\xi \left( a\right) +\xi \left(
a^{\prime }\right) \text{ .}  \label{sumxi}
\end{equation}
\newline
In this representation $S$ is addition.

Similarly, $and$ is represented by a function $P$ such that 
\begin{equation}
\xi \left( ab\right) =P(\xi \left( a\right) ,\xi \left( b\right) ),
\end{equation}
\newline
whenever $a$, $b$, and $ab$ are allowed propositions. The previous argument
can be repeated: the associativity of $and$ constrains $P$ through the
requirement that $\xi (abc)$ be calculable either as $\xi \left( \left(
ab\right) c\right) $ or as $\xi \left( a\left( bc\right) \right) $.
Therefore,

\begin{equation}
P(P\left( u,v\right) ,w)=P(u,P\left( v,w\right) ),  \label{constrp}
\end{equation}
\newline
where $\xi \left( a\right) =u$, $\xi \left( b\right) =v$, and $\xi \left(
c\right) =w$. Our freedom to choose $P$ is further limited, and quite
severely so, by our previous choice of $S$. The distributivity of $and$
relative to $or$ yields an additional constraint: since the number $\xi
(a\left( b\vee c\right) )$ can also be computed as $\xi \left( \left(
ab\right) \vee \left( ac\right) \right) $ consistency requires that 
\begin{equation}
P(u,v+w)=P\left( u,v\right) +P\left( u,w\right) \text{ .}  \label{constrps}
\end{equation}
\newline
Equation (\ref{constrp}) implies that there must exist a function $\eta $,
calculable from $P$, such that 
\begin{equation}
\eta (P\left( u,v\right) )=\eta \left( u\right) +\eta \left( v\right) .
\end{equation}
Taking the exponential of both sides, and letting $\zeta =e^\eta $,
transforms this sum into a product, 
\begin{equation}
\zeta (P\left( u,v\right) )=\zeta \left( u\right) \,\zeta \left( v\right)
\,\,\,\,\text{or}\,\,\,\,\,P\left( u,v\right) =\zeta ^{-1}(\zeta \left(
u\right) \zeta \left( v\right) ).  \label{pzeta}
\end{equation}
\newline
Substitution into eq. (\ref{constrps}), leads to a functional equation for $%
\zeta $, 
\begin{equation}
\zeta ^{-1}(\zeta \left( u\right) \zeta \left( v+w\right) )=\zeta
^{-1}(\zeta \left( u\right) \zeta \left( v\right) )+\zeta ^{-1}(\zeta \left(
u\right) \zeta \left( w\right) ).  \label{zeta}
\end{equation}
\newline
The general solution is $\zeta \left( u\right) =\left( Au\right) ^C$ where $%
A $ and $C$ are constants.$^{\cite{caticha98}}$ Substituting back into eq. (%
\ref{pzeta}) gives $P\left( u,v\right) =Auv$ or $\xi \left( ab\right) =A\xi
\left( a\right) \xi \left( b\right) $. The constant $A$ can be absorbed into
a new function $\psi (a)=A\xi (a)$, so that, finally, the $and$ operation
can be conveniently represented by a simple product rule, \ 
\begin{equation}
\psi \left( ab\right) =\psi \left( a\right) \,\psi \left( b\right) \text{ ,}
\label{prule}
\end{equation}
\newline
while the sum rule remains unaffected, 
\begin{equation}
\psi \left( a\vee a^{\prime }\right) =\psi \left( a\right) +\psi \left(
a^{\prime }\right) \text{ .}  \label{srule}
\end{equation}

To summarize: A quantitative representation of the relations between setups
can be obtained by assigning a number $\psi(a)$ to each setup $a$. Except
for the crucial requirement of consistency there seems to exist a
considerable arbitrariness in the actual choice of $\psi(a)$ but this is
largely illusory: Any consistent assignment is equivalent (i.e., a mere
change of variables) to an assignment where $and$ and $or$ are represented
by the product and the sum rules. Complex numbers assigned in this
particularly convenient way are called ``amplitudes''. This is our main
result.

A first application of these ideas arises from the observation that a single
filter that is totally covered with holes is equivalent to having no filter
at all. Suppose, for simplicity, that the positions of our idealized
particle lie on a discrete lattice, then, using eqs. (\ref{assocor}) and (%
\ref{distrib}), we have (choose $t$ so that $t_i<t<t_f$) 
\begin{equation}
\lbrack x_f,x_i]=\,\,\stackunder{\text{all}\,\vec{x}\,\text{at}\,t}{\vee }%
([x_f,x_t][x_t,x_i])\text{ .}  \label{sigmat}
\end{equation}
\ \newline
Thus motion over a long distance can be analyzed in terms of motion over
shorter steps. The fundamental quantum equation of motion is the equation
for the corresponding amplitudes. Using the sum and product rules, eqs. (\ref
{prule}) and (\ref{srule}), we get 
\begin{equation}
\psi (x_f,x_i)=\sum_{\text{all}\,\vec{x}\,\text{at}\,t}\psi (x_f,x)\,\psi
(x,x_i)\text{ .}  \label{fem}
\end{equation}
\newline
To write it in Schr\"{o}dinger form one introduces,$^{\cite{feynman48}}$ the
notion of a state described by a wave function.

Consider the (not unusual) situation where all reference to the starting
point $(\vec{x}_i,t_i)$, and to those interactions prior to time $t$ can be
ignored. Then the amplitude $\psi (\vec{x},t;\vec{x}_i,t_i)$ can be simply
written as $\Psi (\vec{x},t)$, which we will call the wave function. Writing
eq. (\ref{fem}) as

\begin{equation}  \label{proppsi}
\Psi(\vec{x}_f,t_f)=\sum_{\text{all}\,\vec{x}\,\text{at}\,t}\psi(\vec{x}%
_f,t_f;\vec{x},t)\,\Psi(\vec{x},t)\text{,}
\end{equation}
\newline
shows that $\Psi(\vec{x},t)$ is sufficient to determine future evolution.
The wave function $\Psi(\vec{x},t)$ represents those features of the
particle's history prior to $t$ that are relevant to its evolution after $t$%
; one might say that $\Psi(\vec{x},t)$ represents the state of the particle
at $t$, or perhaps better, that it represents the preparation procedure.

Differentiating eq. (\ref{proppsi}) with respect to $t_f$ and evaluating at $%
t_f=t$ gives the linear Schr\"{o}dinger equation 
\begin{equation}
-i\hbar \frac{\partial \Psi (\vec{x}_f,t)}{\partial t}=\sum_{\text{all}\,%
\vec{x}\,\text{at}\,t}H(\vec{x}_f,\vec{x},t)\,\Psi (\vec{x},t)\text{,}
\end{equation}
\newline
where $H$ is the derivative 
\begin{equation}
\frac{\partial \psi (\vec{x}_f,t^{\prime };\vec{x},t)}{\partial t^{\prime }}%
\Big|_{t^{\prime }=t}\equiv \frac i\hbar \,H(\vec{x}_f,\vec{x},t).
\end{equation}
\newline
Once the assumption is made that the relations among setups are represented
quantitatively in terms of amplitudes, consistency requires that the time
evolution of quantum states be given by a necessarily linear Schr\"{o}dinger
equation. The question of whether non-linear forms of quantum mechanics are
possible should not be addressed at the shallow level of seeking modified
Hamiltonians but should be rephrased at a deeper level as a question about
how relations among physical entities are to be represented by relations
among mathematical ones.

We conclude with a few brief remarks. First, in the quantum logic approach
linearity follows from the reasonable assumption that time evolution is
induced by a symmetry transformation of the proposition system,$^{\cite
{jauch68},\cite{jordan91}}$ which, as a consequence of Wigner's theorem, is
represented by operators that are linear and unitary. No such assumption is
made here.

As introduced here amplitudes play the crucial role of providing a
consistent representation of the relations between various idealized
experimental setups but the important issue of how to use these amplitudes
to predict the outcomes of experiments was not addressed. This gap is filled
in a forthcoming paper$^{\cite{caticha98}}$ where we prove that the
amplitudes must be interpreted according to Born's probability rule.

Finally, we emphasize that this approach yields the standard quantum
mechanics with all its virtues and faults. Important questions such as why
complex numbers instead of other mathematical objects with the required
associativity and distributivity, or those questions associated to the
quantum mechanics of macroscopic objects are left open.

I am indebted to C. Rodriguez, A. Inomata, P. Zambianchi, and J. Kimball for
valuable discussions and many insightful remarks.


\begin{thebibliography}{99}
\bibitem{footnote1}  Since the words `consistency' and `history' both appear
in this letter the reader might be led to suspect a connection with
`consistent histories'. We hasten to point out that, beyond the shared
subject of quantum theory, there is no connection. The 'consistent
histories' approach seeks to clarify issues of interpretation by introducing
rules (the consistency conditions) which restrict discourse to a subset of
histories within which probabilities can be assigned. Our goal is to justify
the formalism itself; we consider propositions to which amplitudes can be
assigned.
\end{thebibliography}
\end{document}